\documentclass[epj]{svjour}
\usepackage{graphics}
\usepackage{color}
\begin{document}
\title{Noise induced transition in Josephson junction with second harmonic}
\author{Nivedita Bhadra\inst{1} 
\thanks{\emph{Present address:} Department of Physical Sciences, Indian Institute of Science Education and Research Kolkata, Mohanpur, Nadia, West Bengal 741246, India}%
}                     
\offprints{}          
\institute{Department of Physical Sciences, Indian Institute of Science Education and Research Kolkata, Mohanpur, Nadia, West Bengal 741246, India }
\date{Received: date / Revised version: date}
%
\abstract{
We show a  noise induced transition in Josephson junction with fundamental as well as second harmonic. A periodically modulated multiplicative colored noise can stabilize an unstable configuration in such a system. The stabilization of the unstable configuration has been captured in the effective potential of the system obtained by integrating out the high frequency components of the noise. This is a classical approach to understand the stability of an unstable configuration due to the presence of such stochasticity in the system and our numerical analysis confirms the  prediction from the analytical calculation. 
\PACS{
      {PACS-key}{discribing text of that key}   \and
      {PACS-key}{discribing text of that key}
     } 
} 
\maketitle
\section{Introduction}
\label{intro}
Presence of noise can induce several interesting phenomena in a nonlinear system, e.g., stochastic resonance\cite{Gammaitoni1998}, dynamical stabilization\cite{Landa1996,Simons2009}, noise induced order\cite{Matsumoto1983}, noise induced chaos\cite{Kautz1985}, broadly known as noise induced transition\cite{VandenBroeck1994a,HorsthemkeW.2006}.We have studied a noise induced transition in a nonlinear system, namely, Josephson junction(JJ) which consists of  first as well as the second harmonic in its current phase relation(CPR).
\par In a Josephson junction, a supercurrent flows between two proximately coupled superconductors through a thin insulating layer due to Cooper pair tunneling. The  value of the supercurrent is proportional to the sine of the difference of the phases $\phi$ of the superconductor order parameter. 
In conventional JJs at equilibrium, the phase difference $\phi$ is zero. In principle, current can be the summation of all harmonics. Generally the single harmonic is adequate for the description of the JJ properties and other higher harmonics are negligible. However, in recent years, there have been a lot of interesting studies on JJ with unconventional current phase relations(CPR)\cite{Goldobin2007,Richard2013,Ouassou2016}.
 ​The second harmonic in the current phase relation has become noticeable in these studies.   Several mechanisms of generation of the second harmonic have also been discussed \cite{Goldobin2007}. For example, in JJs with a ferromagnetic interlayer i.e. S/F/S,  where for some intervals of the exchange field  and F-layer thickness, the ground state corresponds to the phase difference equal to $\pi$, also known as ``$\pi$" junctions. Experimental studies showed a $0-\pi$ transition in S/F/S from the measurements of the temperature dependence of the critical current \cite{Ryazanov2001}. The CPR for JJs is sinusoidal only near
the critical temperature $T_c$,
$I= f_1\sin{\phi}$, $f_1$ being the amplitude of the first harmonic in CPR. At low temperatures, the higher harmonic terms become important and the relation becomes $I=f_1\sin{\phi}+f_2\sin{2\phi}$, where $f_1$ and $f_2$ are amplitudes of the first and the second harmonic of the Josephson current. At the $0-\pi$
transition, the first harmonic becomes zero and the second-harmonic term dominates \cite{Buzdin2005,Sellier2004a,Robinson2006,Robinson2007}. These studies show the higher harmonics to be extremely sensitive to changes in barrier thickness, temperature etc. 
Recently, a robust second harmonic CPR has been established in a Josephson junction of two superconductors separated by a ferromagnetic layer \cite{Pal2014}.
In this work we adopt a classical approach and propose a toy  model for JJ with two harmonics, i.e. $f_2\neq 0$, where a stochastic modulation of the potential barrier is introduced and we have shown that the unstable configuration can be stabilized with this kind of external temporal modulation.\par 
  Several work has been done in the context of JJ with additive Gaussian and non-Gaussian noise. Ref. \cite{Mantegna1998,Dubkov2004,Spagnolo2004} shows how the presence of noise can enhance the stability of fluctuating metastable states. Both Gaussian and non-Gaussian noise sources can affect the escape time of the junctions and the mean lifetime of the metastable states\cite{Augello2010,Valenti2014a}. To the best of our knowledge the effect of multiplicative noise has not drawn much attention. Multiplicative periodic modulation has been discussed in the context of bosonic JJ\cite{Boukobza2010,Sensarma2011,Mann2017}. In experiments the height of the potential
can be modulated externally in Bosonic JJ. Optical lattice
can be a simulator for realizing this kind of modulation
experimentally\cite{Lignier2007}. Depth of the potential, amplitude and frequency of the drive can be varied over a wide range in such a set up. An off-resonant incoherent light source focused at the barrier between them is one of the possible noise source for
such a system\cite{Khodorkovsky2008}. The bosonic JJ can be described by a two mode  Bosonic Hubbard Hamiltonian\cite{Paraoanu2001,Leggett2001,Gati2007}. In \cite{Khodorkovsky2008} theoretical study has been done considering stochastic modulation of the potential barrier in BHH with first harmonic. Ref.\cite{Boukobza2010} shows how the presence of a small amplitude high frequency multiplicative periodic drive leads to the dynamic stabilization of the $\phi=\pi$ configuration. Periodic modulation of
small frequency can easily be implemented in optical set up
by varying the intensity of the counter propagating lasers
forming the optical lattice\cite{Witthaut2008}. Inspired by these works on bosonic JJ we investigated the effect of multiplicative noise or stochastic modulation in JJ with the  first as well as the  second harmonic.
 We are presenting a mathematical model and expect this would in future motivate experiments to introduce this kind of modulation in non bosonic JJ with both harmonic.
  We observe that the unstable state for this JJ model can be stabilized in presence of such stochastic modulation. Power spectrum of the modulating function  has a peak at a high frequency. Due to the presence of this high frequency ``separation of variable" method is applicable. This method was first adopted to understand the dynamic stabilization of the inverted position of a  pendulum periodically driven at the point of suspension\cite{Butikov2001}. Later studies have shown dynamical stabilization of this kind  in presence of noisy drive\cite{Landa1996}. The probability distribution shows that the presence of additive noise cannot induce stabilization of the $\phi=\pi$ position but multiplicative noise can induce such stabilization\cite{Simons2009}. We have considered the presence of a multiplicative modulation of the potential which has the form of a periodically modulated multiplicative colored noise. The idea is to separate the ``fast" and ``slow" variable and integrating out the ``fast" variable. An effective potential $V_{\mbox{eff}}$ for the ``slow" component can be obtained in this way. This elimination of rapid components produces  higher harmonic terms in $V_{\mbox{eff}}$ whose coefficient can be adjusted to generate new local minima at $\phi=\pi$. We adopted this method of  separation of variable to obtain the effective potential for the system. The stable phase for the system has been  achieved from this  effective potential.
\section{Model}
\label{sec:1}
We consider a Josephson junction model with first as well as second harmonic in the current phase relation(CPR). The governing equation for the system is 
\begin{eqnarray}
\label{eq:JJeqn}
\frac{\hbar}{2e}C\ddot{\phi}+\frac{\hbar}{2eR}\dot{\phi}+(I_1\sin{\phi}+I_2\sin{2\phi})=I,
\end{eqnarray}
where $R$, $I_1,I_2$ and $I$ are the junction resistance, the first and second harmonic of Josephson critical current and the bias current; $\phi$ and $C$
represent the phase difference and the capacitance between the two superconductors respectively.  We would consider the case when the  bias current is $I=0$. Rewriting Eq.\ref{eq:JJeqn} we obtain
\begin{eqnarray}
\label{eq:JJwithsecharmonic}
&\ddot{\phi}+2\beta \dot{\phi}+\
 (f_1\;\sin{\phi}+f_2\; \sin{2\phi})=0,
\end{eqnarray}
where $f_1=\frac{2e}{\hbar C} I_1$ and $f_2=\frac{2e}{\hbar C} I_2; \beta=\frac{1}{CR}$ is the damping parameter of Josephson current. For algebraic simplicity, we consider the case $\beta=0$. However, the general case can be considered in the same way (see Eq. \ref{eq:Gwithbeta}). Hence, We can write Eq.\ref{eq:JJwithsecharmonic} in terms of Josephson potential $V$ as

\begin{eqnarray}
\ddot{\phi}+\frac{\partial V}{\partial \phi}=0,
\end{eqnarray}
where
\begin{eqnarray}
\label{eq:Veffwithoutnoise}
V= -f_1\; \cos{\phi}-\frac{f_2}{2}\;\cos{2{\phi}}.
\end{eqnarray}

The condition for stabilization of $\phi=\pi$ position is obtained by putting $\frac{\partial^2 V}{\partial \phi^2}|_{\phi=\pi}>0$, i.e.,
$2f_2>f_1$. Our study is in the other regime, i.e., $f_1>2f_2$. It can be observed from the potential(Eq.\ref{eq:Veffwithoutnoise}) that $f_1>2f_2$ configuration for the system is unstable at $\phi=\pi$. This work  shows how this unstable configuration can be stabilized in presence of modulation of the potential barrier.
 The equation governing the stochastically modulated system is 
\begin{eqnarray}
 \label{eq:JJeqnwithnoise}&\ddot{\phi}+(1+\xi(t))\
 (f_1\;\sin{\phi}+f_2\; \sin{2\phi})=0,
\end{eqnarray}
where $\xi(t)$ is the noise whose power spectrum has a peak at a high frequency compared to the natural frequency of the system when $\xi(t)$ is absent. This choice of $\xi$ ensures that the deviations of
the variable $\phi$, caused by the stochastic vibration, are small. Hence, a separation of ``slow" and ``fast"
variable is possible.
The noise is taken to be Gaussian distributed with zero
mean, $\langle\xi(t)\rangle=0$, and the temporal correlation goes as
\begin{eqnarray}
\label{colorednoisecorr}
\langle \xi(t)\xi(t+\tau)\rangle=D(\tau)=\sigma^2 e^{-\alpha \tau}\cos(\nu \tau).
\end{eqnarray}

Henceforth, angular bracket $\langle...\rangle$ will denote averaging
over the time, where  $\sigma^2$ is the variance of the stochastic process, $\nu$
is the modulation frequency which is taken to be very high compared to the natural frequency of JJ with both harmonic and $\alpha$ is the attenuation parameter. We  have studied the problem under the
limit $1<\alpha \ll \nu$. For the simplicity of calculation, we have not included any additive noise. However, we have shown in the numerical analysis that inclusion of small additive noise does not affect the results significantly. 
\section{Calculation of effective potential $V_{\mbox{eff}}$}
In this section, we would calculate the effective potential for this system. This effective potential approach  was for the first time implemented in the context of dynamic stabilization of a pivot driven pendulum. Dynamical stabilization of $\phi=\pi$ i.e., the inverted position of the system was theoretically explained on the basis of this effective potential. The idea is to obtain the effective potential by splitting the motion into ``fast" and ``slow" variable and averaging over the ``fast" component. In such cases, new harmonics appear in the effective potential. Now the stable phases  can be obtained by finding out the extrema of the effective potential. We adopt the same method for this system.\par
The variable governing the dynamics of this Josephson junction is the phase difference of the superconducting order parameters across the junction $\phi$. We can separate it into ``fast" $\phi_f$ and ``slow" part $\phi_s$. Writing $\phi=\phi_s+\phi_f$, where $\phi_s\gg\phi_f$, Eq.\ref{eq:JJeqnwithnoise} becomes
\begin{eqnarray}
\label{eqntotal}
\ddot{\phi_s}+\ddot{\phi_f}&=&-(1+\xi(t))\;(f_1(\sin{\phi_s}+\phi_f\;\cos{\phi_s})\nonumber \\ & &+f_2(\sin{2\phi_s}+2\phi_f \;\cos{2\phi_s})),
\end{eqnarray}
by keeping terms up to $\mathcal{O}(\phi_f)$. Eq.\ref{eqntotal} involves both the fluctuating part(fast) and the “smooth”(slow) part, each of
which should have its own separate equation of motion.
 Retaining terms linear in $\phi_f$ or $\xi$, and replacing $\phi_s$ by the
noise averaged value $\langle \phi \rangle$ for the fluctuation part, the equation of motion for the fluctuating part is given by

\begin{eqnarray}
\label{fluctuatingpart}
&\ddot{\phi_f}+(f_1\cos\langle\phi\rangle+2f_2\cos{2\langle\phi\rangle})\phi_f= \\& -\xi(t)(f_1\sin{\langle\phi\rangle}+f_2\sin{2\langle\phi\rangle}),
\end{eqnarray}
and that of the ``slow" part is
\begin{eqnarray}
\label{slowpart}
\ddot{\langle\phi\rangle}&=&-(f_1\sin{\langle\phi\rangle}+f_2\sin{2\langle\phi\rangle})-\langle\xi(t)\phi_f\rangle \nonumber  \\ & &  \times(f_1\cos{\langle\phi\rangle}+2f_2\cos{2\langle\phi\rangle}).
\end{eqnarray}

The natural initial condition for $\delta\phi$ is to set
$\delta\phi=\delta{\dot{\phi}}=0$ at the initial time which we may
choose as $t=-\infty$.  With this initial condition, Eq.(\ref{fluctuatingpart}) admits $\langle\delta\phi(t)\rangle=0$ for all
time, as expected.\par

Note that the cross-correlation, as is shown below, is
$\mathcal{O}(1)$, and it contributes to the average motion of the
pendulum.
We can consider $\cos\langle \phi \rangle\approx -1$ as we  are interested in the regime $\phi\approx\pi$.
We would solve Eq. \ref{fluctuatingpart} by Green function $G$.
 The effect of noise can be cast into the form of a  unit force. Eq. \ref{fluctuatingpart} can be written as
\begin{eqnarray}
\label{fluctuationeqn}
\ddot{\phi_f}-(f_1-2f_2)\phi_f=\delta(t-t'),
\end{eqnarray}
where $\delta(t-t')$ is unit force at $t=t'$.
In terms of Green function Eq.\ref{fluctuationeqn} becomes
\begin{eqnarray}
\label{eq:Gforfastpart}
\ddot{G}-\gamma^2 G=\delta(t-t'),
\end{eqnarray}
where $G$ satisfies the initial condition for $t<t'$ with $G$ continuous at $t=t'$ and $\gamma^2=f_1-2f_2$. In this case, the solution $G$ takes the following form
\begin{eqnarray}
\label{eq:Gwithoutbeta}
G(t,t')=\frac{1}{2\gamma}\Big(e^{\gamma(t-t')}-e^{-\gamma(t-t')}\Big).
\end{eqnarray}
Substituting this solution in  Eq.\ref{slowpart}, we find
\begin{eqnarray}
\label{effectiveeqn}
\ddot{\langle\phi\rangle}&=&-(f_1\sin{\langle\phi\rangle}+f_2\sin{2\langle\phi\rangle})-\\ \nonumber  && \frac{\sigma^2}{\nu^2}(f_1\sin{\langle\phi\rangle}+f_2\sin{2\langle\phi\rangle})(f_1\cos{\langle\phi\rangle}+2f_2\cos{2\langle\phi\rangle}),
\end{eqnarray}
 From now on we would opt out $\langle ..\rangle$ for the simplification of symbol. In terms of effective potential $V_{\mbox{eff}}$ we can write
\begin{eqnarray}
\ddot{\phi}+\frac{\partial V_{\mbox{eff}}}{\partial \phi}=0,
\end{eqnarray}
where
\begin{eqnarray}
\label{eq:Veff}
V_{\mbox{eff}}&=&-\Big( f_1-\frac{ f_1 f_2}{2}\frac{\sigma^2}{\nu^2}
\Big)\cos{\phi}-\Big( \frac{f_2}{2}+\frac{f_1^2}{4}\frac{\sigma^2}{\nu^2}
\Big)\cos{2\phi}\nonumber \\&&
 -\frac{f_1 f_2}{2}\frac{\sigma^2}{\nu^2}\cos{3\phi}-\frac{f_2^2}{4}\frac{\sigma^2}{\nu^2}\cos{4\phi}.
\end{eqnarray}
Now, to find the condition for stability of $\phi=\pi$, we simply put the condition $\frac{\partial^2 V}{\partial \phi^2}|_{\phi=\pi}>0$. It gives the condition as
\begin{eqnarray}
\label{stabilitycondition}
 \frac{\sigma^2}{\nu^2}> \frac{1}{(f_1-2f_2)}.
\end{eqnarray}
The stability condition at $\phi=\pi$ shows a divergence  at $f_1=2f_2$. 
\par
Now we study the stable and unstable phases after expanding $V_{\mbox{eff}}$ around $\phi=\pi$. Expanding around $\phi=\pi$ we obtain the effective potential in the following form
\begin{eqnarray}
\label{LZexpansion}
V(\phi)&=&V_0+\frac{1}{2}(2f_2-{f_1})\eta^2+\frac{1}{2!}\frac{\sigma^2}{\nu^2}(f_1-2f_2)^2 \eta^2
+ \nonumber\\&& \frac{1}{4!}(f_1-8f_2)) \eta^4
-\frac{1}{3!}\frac{\sigma^2}{\nu^2}((2f_2-f_1)(8f_2-f_1)) \eta^4 \nonumber\\&& +\frac{1}{6!}(-f_1+32f_2)
\eta^6+\nonumber \\ &&\frac{1}{6!}\frac{\sigma^2}{\nu^2}(16f_1^2-364f_1f_2+1024f_2^2) \eta^6,
\end{eqnarray}
where $\eta=\phi-\pi,\; V_0=f_1-\frac{f_2}{2}-\frac{f_1^2\sigma^2}{4\nu^2}-\frac{f_2^2\sigma^2}{4\nu^2}$. We ignore higher order terms since the highest order term considered here($\eta^6$) is always positive for the range of values we are interested in $f_1<8f_2$.\par
Let us now consider a few special situations.
\begin{enumerate}
\item In the modulation i.e. $\frac{\sigma^2}{\nu^2}=0$ and $f_1=2f_2$, the series(Eq.\ref{LZexpansion}) starts with the
  quartic term, with the coefficient of $\eta^4$ as negative $(-\frac{f_2}{4})$.  The
  coefficient of $\eta^6$ is positive ($\frac{1}{24}f_2$), providing stability to the system. Therefore, ${\partial^2 V_{\mbox{eff}}}/{\partial
    \phi^2}|_{\pi}<0$. Hence, $\eta=0 \; \mbox{or} \; \phi=\pi$ is an unstable configuration for the system.
   
\item In presence of noise .i.e.,when, $\frac{\sigma^2}{\nu^2}\neq 0$, and the condition $\frac{\sigma^2}{\nu^2}>\frac{1}{(f_1-2f_2)}$ is satisfied, the coefficient of $\eta^2$ is positive. In this case, ${\partial^2 V_{\mbox{eff}}}/{\partial
    \phi^2}|_{\pi}>0$, hence, $\eta=0$ is stable configuration for the system.
\item At $f_1=2f_2$, both  $\eta^2, \; \eta^4$
terms vanish. Effective potential becomes
\begin{eqnarray}
V_{\mbox{eff}}=\frac{3}{2}f_2-\frac{1}{4}f_2
\eta^4+\frac{1}{4!}f_2\eta^6.
\end{eqnarray}

 Hence, it is an  unstable configuration at $\phi=\pi$.
 \end{enumerate}
When $\beta\neq 0$, the governing equation becomes
\begin{eqnarray}
 \label{eq:JJeqnwithnoisewithbeta}&\ddot{\phi}+2\beta \dot{\phi}+(1+\xi(t))\ (f_1\;\sin{\phi}+f_2\; \sin{2\phi})=0.
\end{eqnarray}
The Green function solution is 
\begin{eqnarray}
\label{eq:Gwithbeta}
G(t,t')&=&\frac{1}{2\sqrt{\gamma^2+\beta^2}}\Big( \exp{(-\beta+\sqrt{\gamma^2+\beta^2})(t-t')}\nonumber \\ && -\exp{(-\beta-\sqrt{\gamma^2+\beta^2})(t-t')} \Big).
\end{eqnarray}
If $\beta=0$, Eq.\ref{eq:Gwithbeta} reduces to Eq.\ref{eq:Gwithoutbeta}. Details of the analytical expression for $\beta\neq 0$ is not given. However, in the numerical analysis part we have included a small value of damping parameter $\beta$.
\section{Numerical results}
The theoretical results we have obtained in the previous section are approximate.
To observe the shape of the time evolution for phase fluctuation $\phi$ we have performed some numerical analysis. \par
\begin{figure}
\resizebox{0.5\textwidth}{!}{%
  \includegraphics{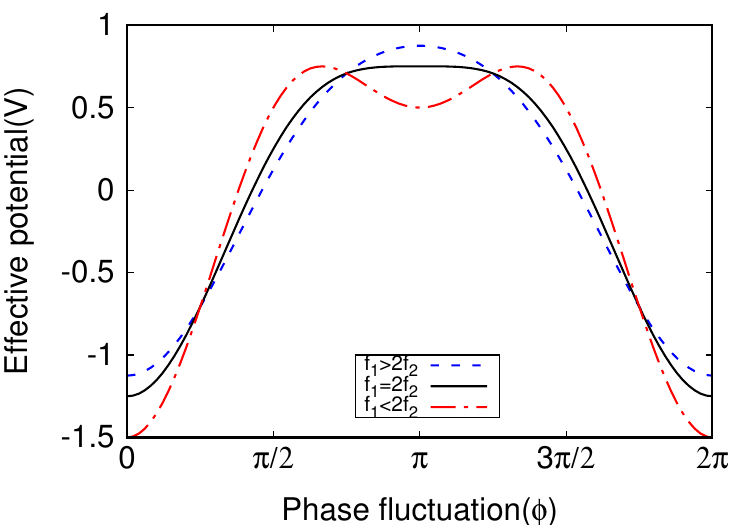}
}
\caption{Plots of Josephson potential(Eq.\ref{eq:Veffwithoutnoise}) in absence of stochastic modulation $\xi(t)$. The solid line shows the potential curve for $f_1=2f_2$. The dashed  and dash-dotted curve shows the potential for $f_1>2f_2$ and $f_1<2f_2$ configuration respectively. For $f_1>2f_2$, $\phi=\pi$ is a maximum in the potential(unstable configuration) whereas, for $f_1<2f_2$, the system achieves a minimum in the potential(stable configuration). $\phi=0$ is always a stable configuration for the system.}
\label{fig:veffwithoutnoise}       
\end{figure}
\begin{figure}[h!]
\resizebox{0.5\textwidth}{!}{
\includegraphics{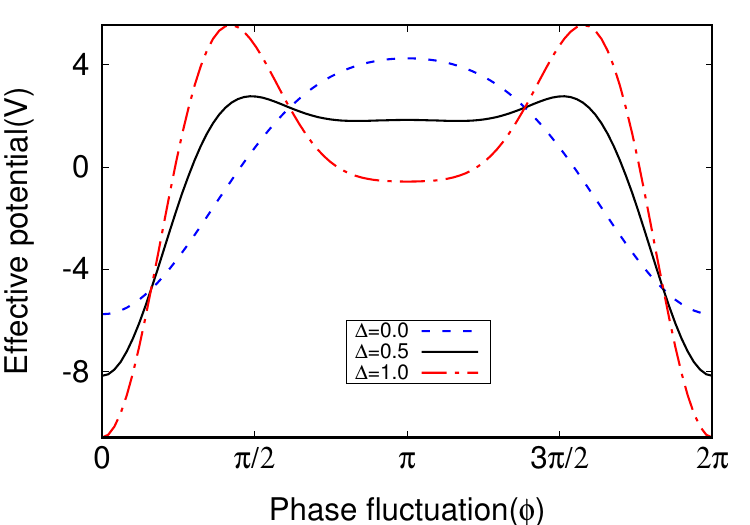}
}
\caption{Plots of effective potential in presence of modulation $\xi(t)$(Eq.\ref{eq:Veff}). These plots are drawn for $f_1=2,f_2=0.5$ ($f_1>2f_2$).$\Delta$ is defined as $\frac{\sigma^2}{\nu^2}$, the parameter for the stochastic modulation $\xi(t)$. The dashed curve is for the potential in $f_1>2f_2$ regime in modulation i.e., $\Delta=0$. It shows $\phi=\pi$ is a maximum(unstable configuration). Effective potential with the parameter $\Delta=0.5$ and $1$ is shown in the solid curve and dash-dotted curve respectively. Gradual increase of the parameter $\Delta$ shows that $\phi=\pi$ gradually becomes a minimum of the potential(stable configuration).}
\label{fig:veffwithnoise}
\end{figure}

According to our analytical calculation for the potential (Eq.\ref{eq:Veffwithoutnoise}) of the system we have found that $\phi=\pi$ is an unstable configuration in the regime $f_1>2f_2$. Fig.\ref{fig:veffwithoutnoise} shows effective potential in absence of any modulation of the potential $\xi(t)$.
The solid line shows the shape of the potential for    $f_1=2f_2$. The plot shows a maximum for $f_1>2f_2$ whereas, in the regime $f_1<2f_2$, it shows a minimum at $\phi=\pi$. Gradually curvature of the potential changes  from positive to negative undergoing a zero value  indicating a transition  at $f_1=2f_2$.
In order to find the effect of modulation $\xi(t)$, we plot in Fig.\ref{fig:veffwithnoise} the effective potential obtained from the analytical calculation(\ref{eq:Veff}). The plot shows $\phi=\pi$ is a maximum potential configuration in absence of $\xi(t)$ i.e.,$\frac{\sigma^2}{\nu^2}=0$, whereas, for nonzero values of noise it gradually becomes a minimum at $\phi=\pi$.

\begin{figure}
\resizebox{0.5\textwidth}{!}{
\includegraphics{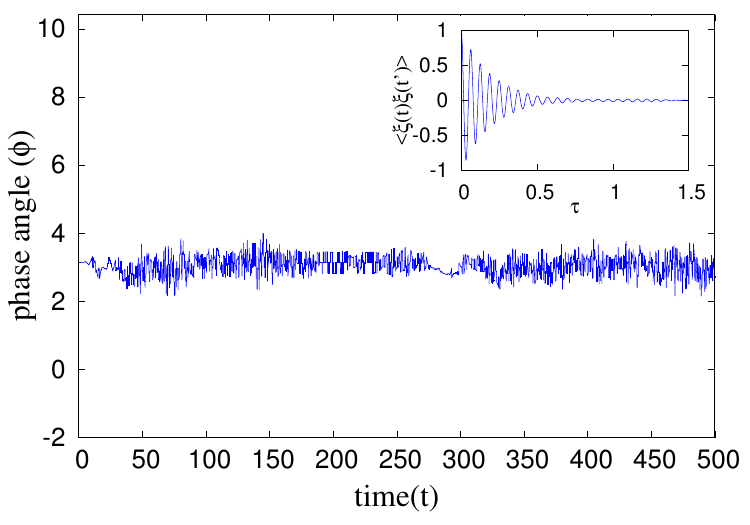}
}
\caption{ The main panel shows time series of the phase fluctuation $\phi$. We have numerically simulated Eq.\ref{eq:JJeqnwithnoisewithbeta}.Parameters are $\alpha=1.5,\beta=0.3,\nu=10$. The inset figure shows autocorrelation of the periodically modulated colored noise $\xi(t)$.}
\label{fig:noisePS}
\end{figure}

\begin{figure}
\resizebox{0.5\textwidth}{!}{
\includegraphics{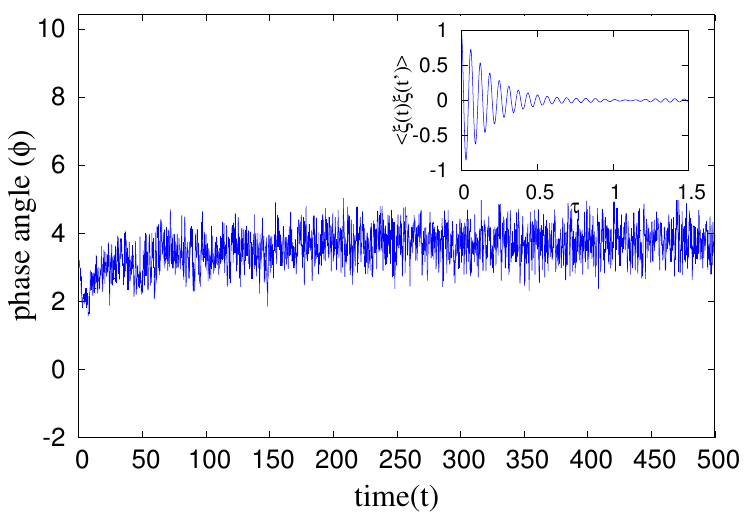}
}
\caption{ The main panel shows time series of the phase fluctuation $\phi$. We have numerically simulated Eq.\ref{eq:JJeqnwithnoisewithbeta} including small magnitude of additive noise.Parameters are $\alpha=1.5,\beta=0.3,\nu=10$. The inset figure shows autocorrelation of the periodically modulated colored noise $\xi(t)$.}
\label{fig:noisyphi_additive}
\end{figure}

To observe the effect of colored noise on the time evolution of $\phi$ we have simulated Eq.\ref{eq:JJeqnwithnoisewithbeta} by Euler algorithm. We have taken the parameter values as, $\beta=0.3, \alpha=1.5,\nu=10,f_1=2,f_2=0.5 (f_1>2f_2)$. Time step was chosen to be $0.001$. We have taken $\xi(t)$ as sufficiently wide band noise. Fig.\ref{fig:PSofcolorednoise} shows the power spectral density of $\xi(t)$.
The spectrum shows a sharp peak at the frequency $\nu=10$. $\xi(t)$ is generated by a forced oscillator with frequency $\nu$ which has a damping $\alpha=1.5$. The applied force is a delta correlated random force. The solution of this randomly forced damped oscillator is fed into the JJ equation(Eq.\ref{eq:JJeqnwithnoisewithbeta}). The variance is $\sigma^2\approx 100.3$ such that $\frac{\sigma^2}{\nu^2} >\frac{1}{f_1-2f_2}$ which satisfies the stability condition at $\phi=\pi$. The initial condition for $\phi$ has been kept $\pi+0.001$ as we are interested in observing the motion of $\phi$ around $\pi$. $\dot{\phi}$ has been kept as  $0.01$. The potential shown in Fig.\ref{fig:veffwithnoise} is also consistent with the nature of the trajectory for phase fluctuation $\phi$ in Fig.\ref{fig:noisePS}. In absence of $\xi(t)$, the potential shows only one minimum at $\phi=\pi$, whereas, in presence of $\xi(t)$ another minimum appears at $\phi=\pi$ in the effective potential. All these plots shown are an average of 40 different realizations.
In Fig.\ref{fig:noisyphi_additive} we show the timeseries of $\phi$ in presence of additive white noise. The equation governing such case would be
\begin{eqnarray}
\ddot{\phi}+(1+\xi(t))(f_1\sin{\phi}+f_2\sin{2\phi})+\eta(t)=0\nonumber,
\end{eqnarray}
where $\langle\eta(t)\eta(t')\rangle=\delta(t-t')$. This kind of additive white noise might be present due to thermal fluctuation(effect of environment) in such a system. The amplitude of the thermal noise we consider is $\mathcal{O}(10^{-1})$. We have not studied the large thermal fluctuation regime. From Fig. \ref{fig:noisyphi_additive} we observe that nature of the timeseries of $\phi$ does not change noticeably in this regime.

\section{Conclusion}
In this work, we have shown how the modulation $\xi
(t)$ having the correlation of a periodically modulated colored noise can stabilize unstable configuration in Josephson junction with the first and the second harmonic. There is a surge in interest in the study of JJ with dominant second harmonic term and its several ground states recently. We take a classical route to understand the $\phi=\pi$ state. Our study shows appearance of this state due to the presence of such multiplicative noisy modulation in the system. This kind of noise can be implemented in Bosonic JJ as discussed in Sec.\ref{intro}.
The stable state has been detected by analyzing the effective potential we obtained.This approximate analysis shows how new harmonic appears in the effective potential. We also observe the stability of $\phi=\pi$ in the numerical analysis of the stochastic equation governing the system.
In general for a JJ system with first harmonic, the phase fluctuation for ground state is $\phi=0$. In presence of second harmonic, it depends on the amplitude of the harmonic of the potential barrier. In our current study, we have shown how the $\phi=\pi$ can be made a ground state in presence of a stochastic modulation. In JJ, the value of phase fluctuation, i.e. $\phi$ is actually a measure of Josephson current. Since the Josephson current is proportional to the  sine argument of the phase fluctuation, $\phi$ being negative indicates inversion of the direction of the current. The condition to make the $\phi=\pi$ as a ground state depends on the parameter of the applied modulation of the potential. 
\section{Acknowledgements}
The author would like to thank Siddhartha Lal and Anandamohan Ghosh
for  valuable discussions.
\section{Authors contributions}
Both the analytical calculations and the numerical analysis have been performed by N.B.

%

\end{document}